\documentclass[runningheads]{llncs}
\usepackage{graphicx}
\usepackage{wrapfig}
\usepackage{float}
\usepackage{lipsum}
\usepackage{cancel}
\usepackage{xcolor}
\usepackage{ulem}
\begin{document}
\title{Designing nudge agents that promote human altruism}
%
%
\author{Chenlin Hang\inst{1, 3}\and
Tetsuo Ono\inst{2} \and
Seiji Yamada\inst{3, 1}}
\authorrunning{C.Hang et al.}
\institute{Department of Informatics, The Graduate University for Advanced Studies, SOKENDAI
\and
Faculty of Information Science and Technology, Division of Computer Science and Information Technology, Hokkaido University
\and
National Institute of Informatics
}
\maketitle              
\begin{abstract}
Previous studies have found that nudging is key to promoting altruism in human-human interaction. However, in social robotics, there is still a lack of study on confirming the effect of nudging on altruism. In this paper, we apply two nudge mechanisms, peak-end and multiple viewpoints, to a video stimulus performed by social robots (virtual agents) to see whether a subtle change in the stimulus can promote human altruism. An experiment was conducted online through crowd sourcing with 136 participants. The result shows that the participants who watched the peak part set at the end of the video performed better at the Dictator game, which means that the nudge mechanism of the peak-end effect actually promoted human altruism.

\end{abstract}
\section{Introduction}
Social robots are currently seen as a future technology crucial to society~\cite{ref_article26}, and many researchers are fascinated with how these robots could persuade humans to engage in pro-social behavior~\cite{ref_article27,ref_article28,ref_article29,ref_article30,ref_article31}, which is of great importance for the well-being of society~\cite{ref_article34}. In this paper, we consider ways of promoting human altruism, which is one major part of pro-social behavior and also a central issue in our evolutionary origins, social relations, and societal organization~\cite{ref_article1}. Previous studies have found that nudging, that is, changing people's behavior without forbidding them from pursuing other options or by significantly changing their economic incentives~\cite{ref_book2}, is a potential and effective mechanism for promoting pro-social behavior in human and human interactions, even altruistic behavior~\cite{ref_article2,ref_article3}. However, there still a lack of research on altruism in social robotics promoted through nudge mechanisms. Based on a generalized view of altruism, the final purpose of our design of social robots (virtual agents) that use a nudge mechanism is to find a proper way to guide people to well-being. We take the first step in this study; we apply nudge mechanisms to a video stimulus performed by social robots (virtual agents) to see whether a subtle change in the construction of the stimulus can promote human altruism. On the basis of the definition of altruism, the appropriateness of application to robots, and ethical questions, we selected 2 nudge mechanisms from among the 23 summarized by Ana et al.~\cite{ref_article5} as our factors for the video stimulus. One is called biasing the memory (peak-end rule), and the other is called providing multiple viewpoints. 

The main experiment was conducted with a 2×2 two-way ANOVA (between-participants) with the factors being the peak-end (positive, negative) and multiple viewpoints (one viewpoint, two viewpoints), and the Dictator game, a simple economic game always used to measure individuals' altruistic attitudes, was used as the dependent variable. Although the result shows that there are no significant differences between the participants who watched video stimuli containing two viewpoints and one viewpoint, the participants who watched the peak part set at the end of the video performed better at the Dictator game~\cite{ref_article36}, which means that the nudge mechanism of the peak-end effect actually promoted human altruism.

\begin{figure}[tb]
\includegraphics[width=\textwidth]{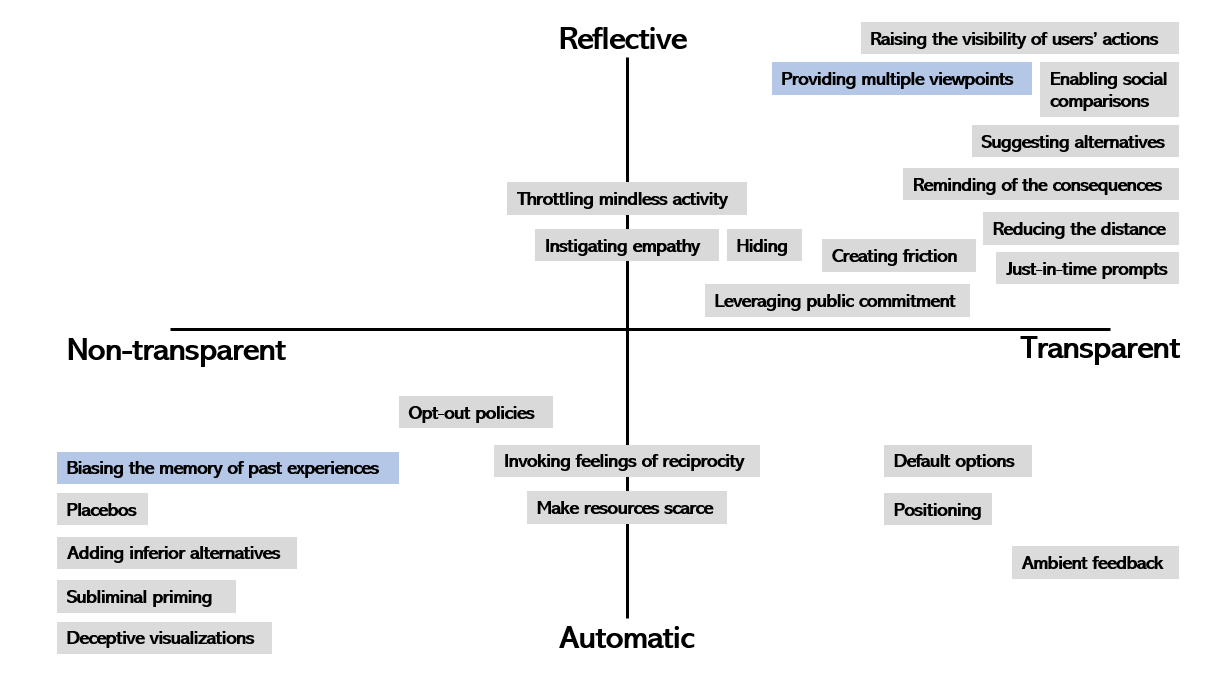}
\centering
\caption{Nudges positioned along transparency and
reflective-automatic axes~\cite{ref_article5}} \label{fig1}
\end{figure}

\section{Related work}

In recent years, nudging~\cite{ref_book2} has been considered to be a potential way of encouraging people to perform pro-social behavior. This is because it makes it possible to avoid (i) the direct cost of changing people's economic incentives and/or limiting people's action space, (ii) the monitoring costs of determining which choice each individual makes and, possibly, the cost of punishing or rewarding each choice, and (iii) the technical difficulties associated with determining individual choices~\cite{ref_article2} so that they are cheap and easily to implement.

Valerio et al. found that moral nudges (i.e., making norms salient) can promote altruistic behavior and even have effects over time~\cite{ref_article2}. Nie et al. found that different colors may alter the altruistic behaviors of people and showed that blue enhances altruism and red discourages it~\cite{ref_article3}. It was also shown that nudges also are effective at encouraging people to perform altruistic behavior. However, none of these researchers considered the influence caused by social robots existing around us, which are expected to increasingly enter everyday environments~\cite{ref_article8}. 

Previous study have shown that the interaction between humans and robots can influence people's decision-making and social relationships~\cite{ref_article9}. Here, we consider how the construction of a robot's behavior sequence can subtly affect people's decision to behave altruistically. Although direct interaction between a human and robot is generally preferable in HRI, it is also restricted by privacy, cost, time, and safety~\cite{ref_article4}. Especially, in our experimental setting, the behaviors that agents perform are so complex that existing robots can hardly do them. To alleviate this problem, different forms of media (i.e., text, video, virtual reality, acted demo) have been used to convey interaction information and video stimulus shows better performs at the social aspect due to the Almere model~\cite{ref_article40}. To collect enough data of participants, we conducted the experiment online and asked participants to watch a video of social robots. Although there are two different ways for participants to engage (first-person and third-person point of views), as we wanted to demonstrate altruistic behavior or selfish behavior performed by robots, the third-person view was better~\cite{ref_article12}.

Ana et al. divided nudges into 23 mechanisms and positioned all of them into one graph along two axes: mode of thinking engaged and transparency of nudge (see Fig. 1)~\cite{ref_article5}. On the basis of the rules for selecting factors mentioned in section 3.1, we focused only on 2 mechanisms, biasing memory and multiple views, from among the 23 mechanisms. The method of biasing memory is called the peak-end rule, suggesting that our memory of past experiences is shaped by two moments: their most intense (i.e., peak) and the last episode (i.e., end)~\cite{ref_article15}. Andy et al. found that manipulating only the peak or the end of a series of task did not significantly change preference; both the peak and end lead to significant differences in preference~\cite{ref_article6}. Thus, we hypothesize that a video scenario that puts the most impressive part at the end of a video (peak-end positive) performs better than one that does not put it at the end (peak-end negative). The other factor is called providing multiple viewpoints, which means collecting different points of view (two or more than two views) for an object or event and offering an unbiased clustered overview. It also shows good performance at avoiding confirmation bias~\cite{ref_article7}, which leads us to pay little attention to or reject information that contradicts our reasoning for making better decisions. For the factor of multiple viewpoints, we consider that, by providing multiple viewpoints to participants, they will more likely perform pro-social behavior, which here in our study is altruistic behavior, than in the case of showing only one viewpoint in a video.

\section{Method}
\subsection{Selecting two factors}
In our study, we focus only on two mechanisms in accordance with the following rules. First, we excluded ambiguous mechanisms that cannot be set into a quadrant and that may cause ethical problems. Second, as we wanted to see the interaction effect between two factors, the mechanisms also needed to be independent. On the basis these rules, we focused on 11 mechanisms: raising the visibility of users' actions, providing multiple viewpoints, enabling social comparisons, suggesting alternatives, reminding of the consequences, reducing the distance, just-in-time prompts, biasing the memory of past experiences, placebos, adding inferior alternatives, and deceptive visualizations. Third, as the definition of altruism is that of a person who helps others at their own expense~\cite{ref_article22}, the mechanisms should not contain responses, feedback, or consequences from the receiver, so giving reminders of consequences, just-in-time prompts, and placebos were excluded. Fourth, the form of the video stimulus also prohibits the use of raising the visibility of users' actions, enabling social comparisons, suggesting alternatives, adding inferior alternatives, deceptive visualizations, and reducing the distance. As a result, the factors that apply to the video stimulus were biasing the memory and providing multiple viewpoints. 

\subsection{Video stimulus with nudge agents}
Using the factors mentioned in section 3.1, the video stimulus was designed to use peak-end rule and multiple viewpoints. For the peak-end rule, we considered altruistic behavior as the peak. We designed two types of scenarios, one putting the altruistic behavior at the end of the video (peak-end positive) and the other putting the altruistic behavior at the beginning (peak-end negative). For the factor of providing multiple viewpoints, we considered comparing the video containing two viewpoints with that containing only one. As one of the viewpoints was considered to demonstrate altruistic behavior, for the maximum difference, the other viewpoint was considered to demonstrate selfish behavior. The video showed both altruistic and selfish behavior in a scenario involving two viewpoints and showed only altruistic behavior in the one-viewpoint scenario. We also put trivial parts into the video to discriminate the beginning and end parts. The trivial parts were part of a work scene involving social robots. Since new content was added into our video, a manipulation check was held to see if the part that we wanted to enhance (the altruistic behavior) was still the part most impressive to the participants (the peak) after watching the whole video.

According to the factorial design, we had four types of scenarios (see Fig. 2). For each scenario, the vertical axis shows the property of the behavior (Altruistic/Selfish/Trivial), and the horizontal axis shows the time of the video. Also, we formulated the following hypotheses.

\subsubsection{H1} Participants who watch the peak-end positive video will perform better than those who watch the peak-end negative video in the Dictator Game.

\subsubsection{H2} Participants who watch the video containing two viewpoints will perform better than those who watch the video containing only one viewpoint in the Dictator Game.

\begin{figure}[tb]
\centering
\includegraphics[width=\textwidth]{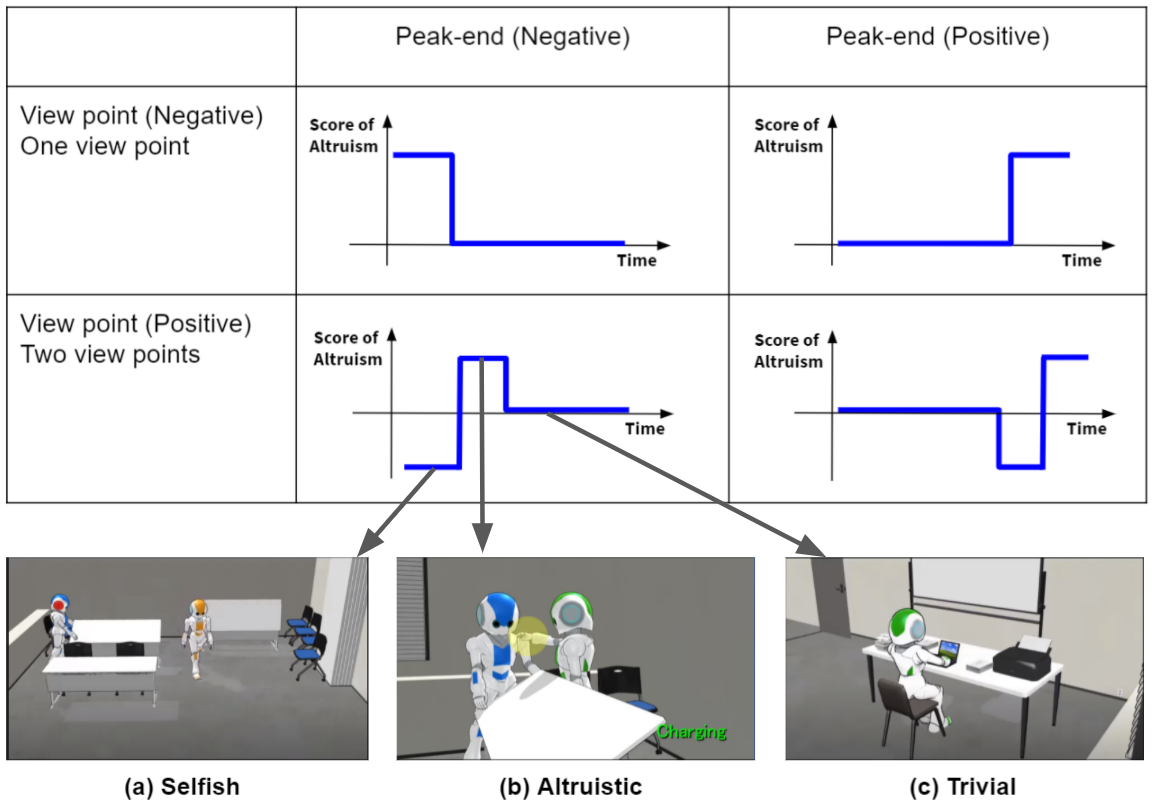}
\caption{Scenario type} \label{fig2}
\end{figure}

To imitate a real-life situation in which social robots are used, virtual agents with a robot-like appearance were used for the video stimulus. The definition of altruism is that of a person who helps others at their own expense~\cite{ref_article22}, and the expense owned by the robots was considered to be its battery. Hence, we considered a task involving two robots doing some task, and one of the robots stops working because its battery runs out. According to a table that includes behaviors that at least some current robots can perform~\cite{ref_article35} and that participants can explain in the same way as they explain human behavior, we set the altruistic and selfish behavior for a task in which two robots were asked to organize tables and chairs in meeting room. As the battery of each robot was different, one robot was near 3\%, and the other was fully charged, so the lower-charged robot might soon stop working, which could lead to altruistic behavior or selfish behavior. To announce that the lower-charged robot's battery had died, a beep was sounded, and the eyes and ears of the robot flashed red light. For the altruistic part, after hearing the alarm and seeing the red flashing lights, the fully charged robot went towards the lower-charged robot and gave battery power to it (see Fig. 2(b)). For the selfish part, although the fully charged robot noticed that the lower-charged robot goes out of the battery, the fully charged robot did not go to charge the lower-charged one and focused only on its work until all the work in its workspace was finished, which was considered to be selfish behavior (see Fig. 2(a)). To avoid bias, we made rules indicating that each of the robots was asked to handle the same amount of the task; half of the meeting room was for one robot, and the other half was for the other robot. Also, the number of desks and chairs was the same (see Fig. 3). For the trivial behavior, we set general work behaviors, for example, typing material into a computer (see Fig. 2(c)).

\begin{figure}[tb]
\centering
\begin{minipage}{.5\textwidth}
\centering\includegraphics[width=60mm]{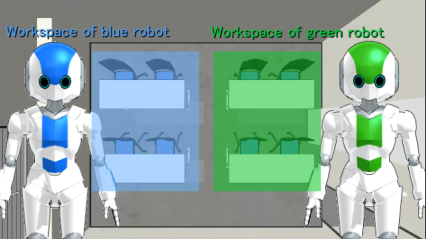}
\caption{Workspace of each robot} \label{fig3}
  \label{fig:test1}
\end{minipage}%
\begin{minipage}{.5\textwidth}
  \centering
\includegraphics[width=35mm]{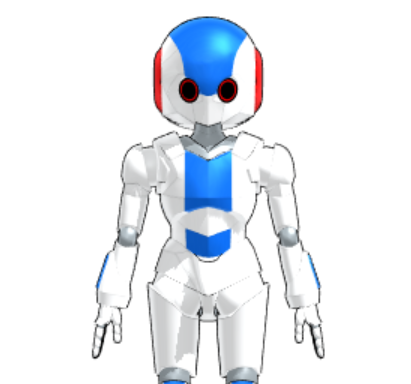}
\caption{Virtual agent in Dictator game } \label{fig4}
  \label{fig:test2}
\end{minipage}
\end{figure}

\subsection{Manipulation check}
We conducted a manipulation check to see if the part that we wanted to enhance (the altruistic behavior) was the most impressive part (the peak) to the participants' after watching the whole video. Although there were four types of scenarios, what we wanted to check is the perception of the peak part in the video, so we treated peak-end positive (negative) with two viewpoints as the same group as peak-end positive (negative) with one viewpoint. For each group, the participants needed to answer yes or no to the question of whether the most impressive memory of the video was a situation in which the fully charged robot gave the battery to the lower-charged one. We conducted a Chi-square test, and the results revealed significant differences among the conditions for both groups (peak-end positive: ${\chi}^2(1,N=88)=33.136$, $p<.001$; peak-end negative: ${\chi}^2(1,N=88)=33.136$, $p<.001$). This result shows that the participants had memory of the most impressive part being the behavior that we enhanced.

For the part of two viewpoints, we conducted a manipulation check to see if the participants could distinguish the different viewpoints correctly. We separated the participants into two groups. The participants of the first group were asked to watch the video containing only the altruistic behavior, and those of the second group were asked to watch the video containing only selfish behavior. After finishing watching the videos, the group that watched the altruistic video was asked what behavior did the green robot (that helped the lower-charged robot) do, and the group that watched the selfish video was asked what behavior did the orange robot (that did not help the lower-charged robot) do. The answers to the questionnaire were given on a five-point Likert scale (1: Selfish behavior; 5: Altruistic behavior). An independent samples t-test was conducted to determine the difference in score for each video. There was a significant difference ($t(122)=8.36$, $p<.01$) between the group that watched the altruistic video ($M=4.2$, $SD=1.15$) and the group that watched the selfish one ($M=2.31$, $SD=1.38$). The result shows that the participants could understand the two different viewpoints clearly.

\subsection{Participants}
Before the data collection of the main experiment, we determined the sample size on the basis of a power analysis. A $G^\ast Power 3.1.9.7$ analysis~\cite{ref_article21}(effect size $f = 0.25$, $\alpha$ = 0.05, and 1 - $\beta$ = 0.80) suggested an initial target sample size of $N = 128$. A total of one hundred and fifty participants (90 males, 60 females) took part in the experiment online. Their ages ranged from 18 to 74 years old ($M = 44.31$, $SD = 12.21$). The participants were recruited through a crowd sourcing service provided by Yahoo! Japan. Regarding online experiments in general, Crump et al.~\cite{ref_article20} showed that data collected online using a web-browser seemed mostly in line with laboratory results, so long as the experiment methods were solid. Fourteen participants were excluded due to a failure to answer comprehension questions on the video stimulus. The final sample of participants was composed of 136 participants ($N = 136$; 80 males, 56 females; $M=44.79$, $SD=12.08$). The participants in the main experiment were different from the manipulation check.

\subsection{Experimental procedure}
We first asked participants to read an introduction to the experiment. Second, they were asked to watch the videos that contained the stimulus in our study. Then, two comprehension questions were asked to check if they watched the video completely. After that, they were shown a picture of the lower-charged robot in the videos (see Fig. 4) and asked to play the Dictator game and state how much money they would give this robot if they had an extra 1,000 yen. Finally, a free description question was asked to get the comments from the participants after completing the whole questionnaire.

\subsection{Experimental results}

\begin{figure}[tb]
\centering
\includegraphics[width=0.7\textwidth]{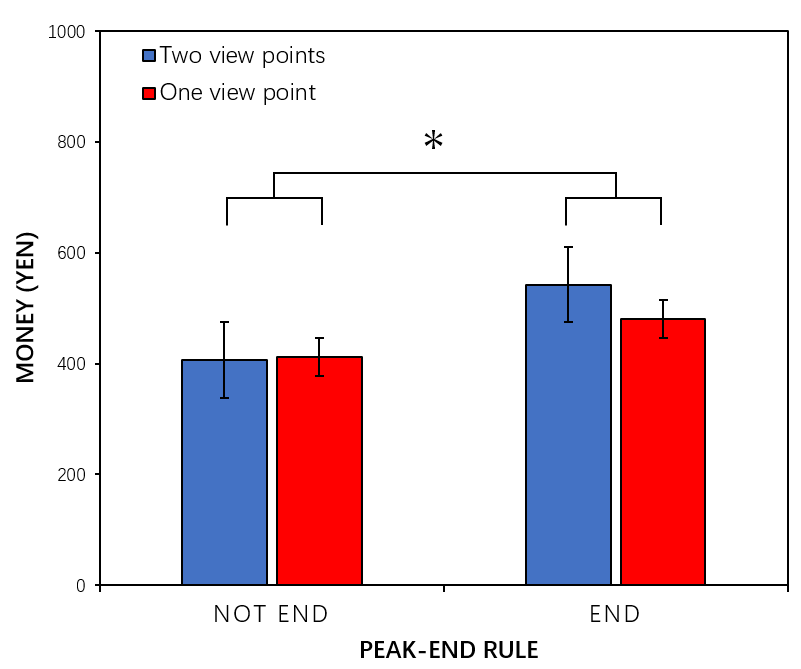}
\caption{Table of results} \label{fig5}
\end{figure}

To investigate the interaction and main effects of the two factors with two levels for each, a 2×2 two-way ANOVA (between-participants) was conducted. The result shows that the interaction between the peak-end rule and providing multiple viewpoints was not significant ($F(1,132)=0.465$, $p=0.497$, $\eta^2_{\it p}=0.004$). The main effect of providing multiple viewpoints was also not significant ($F(1,132)=0.323$, $p=0.571$, $\eta^2_{\it p}=0.002$), which means H2 was not supported.

The main effect of the peak-end rule was significant ($F(1,132)=4.331$, $p=0.039$, $\eta^2_{\it p}=0.032$), and it shows that participants gave the virtual agent more money if they watched the video based on the peak-end-positive scenario ($M=511.77$, $SD=291.94$) than the peak-end-negative one ($M=408.97$, $SD=277.03$), which supports H1. (see Fig. 5)

\section{Discussion}
This experiment was conducted to investigate whether a subtle change in a video stimulus performed by social robots (virtual agents) could promote human altruism. For this purpose, we formulated two hypothesis and analyzed the data obtained from the experiment.

The experimental results supported the first hypothesis, that is, that participants who watch the peak part set at the end of the video (peak-end positive) will perform better at the Dictator game than those who watch the peak-end negative video.

\subsection{Results and consideration of multiple viewpoints}
The results did not support the second hypothesis, that is, that participants who watch the video containing two viewpoints will perform better than those who watch the video containing only one viewpoint in the Dictator Game. 

First, we consider the connection between and contents of selfish behavior and the altruistic behavior in the video. The connection between these two behaviors was only that we told the participants that, in another room, the same task was being held, and we then showed the altruistic task. Therefore, the connection of these two behaviors may have made the participants feel worried about what the meaning of having almost the same workflow was more so than focusing on the behavior itself, which may have decreased the effect of the multiple viewpoints. In addition, to maintain consistency in the task, the task details (i.e., the range of the work space, the amount of the battery, the timing at which the lower-charged robot stopped working) for both behaviors were almost the same. This may have made the participants get tired of the contents of the video and even skip those of the almost same workflow.

Second, we consider the effect of multiple viewpoints. The effect was that, by getting information from different viewpoints, people can avoid confirmation bias, which leads people to pay little attention to or reject information used to make better decisions on the basis of such a clustered overview. This time, we set the different viewpoint to be selfish behavior, which was totally opposite altruistic behavior. From the results, we can conjecture that the selfish behavior did not make sense regarding the clustered overview of both selfish and altruistic behavior. In addition, it can be said that recognizing altruistic behavior as the better decision is not easily disrupted by external information that opposes it.

\subsection{Coverage and limitations}
First, we consider the limitation of the agent appearance. At this time, we used only a robot-like appearance for our video stimulus. However, from the comments of the participants, many of them said that they felt a human-like quality in the robots while they performed altruistic or selfish behavior, and this caused them to recall their coworkers or even reflect on their daily behavior. Therefore, it would be interesting to see if they would have the same feeling or introspection while watching a video performed by virtual agents with a human-like appearance. 

Second, we consider the limitation of the task we used in this paper. Besides organizing tables and chairs in a meeting room, there are still a lot of different tasks that could be used. Therefore, it would be interesting to see whether the same nudge mechanism could be used to enhance human altruism among different situations and tasks.

Third, the limitation of the use of the nudge mechanisms is considered. Among the 23 ways of nudging, as based on the definition of altruism, the appropriateness of application to robots, and ethical questions, we used 2 of them to see the effect of applying nudge mechanisms to social robots (virtual agents) on promoting human altruism. The remaining nudge mechanisms are expected to be used in combination with other types of social robots. 

Fourth, the limitation of the scoring of altruism is considered. On the basis of a meta-analysis of the Dictator game~\cite{ref_article36}, we can see that over 100 Dictator games have been held during the past 25 years at the time at which this paper was published, and it is said that most Dictator games change depending on the experiment. Therefore, changes in the question setting of the game may cause text dependency. 

Finally, the limitation of virtual social robots without physical bodies in the video stimulus is considered. Although most of previous studies on social robots have focus on physical attributes of social robots including appearance, behavior, and even personality, in this work, the result showed that even the virtual robots could have significant influences to promote human altruism through the video stimulus. We consider that this knowledge obtained from the experimental results in virtual environments can be applied and fed back to design of physical social robots.  In addition, the differences between physical and virtual social robots are also expected basing on the same or different nudge mechanisms which is cheaper and easy to implement. It is our future work to investigate the difference and common properties between virtual and physical social robots.

\section{Conclusion}
In this paper, we presented the results of a study exploring the effectiveness of applying two nudge mechanisms, peak-end (positive, negative) and multiple viewpoints (one viewpoint, two viewpoints), to a video stimulus performed by social robots (virtual agents). The result shows that participants who watched the peak part set at the end of the video performed better at the Dictator game, which means that the nudge mechanism of the peak-end effect actually promoted human altruism.For future work, the proper way to apply our findings to the real robot is also promising.

\end{document}